# DO SWITCHING COSTS MAKE MARKETS MORE OR LESS COMPETITIVE?: THE CASE OF 800-NUMBER PORTABILITY*

Abstract

Do switching costs reduce or intensify price competition in markets where firms charge the same price to old and new consumers? Theoretically, the answer could be either "yes" or "no," due to two opposing incentives in firms' pricing decisions. The firm would like to charge a higher price to previous purchasers who are "locked-in" and a lower price to unattached consumers who offer higher future profitability. The net effect relative to a market without switching costs will depend on the mix of old and new consumers and the relative strength of these two effects. I demonstrate this ambiguity in an infinite-horizon theoretical model that, in contrast to previous models, allows for actual switching in equilibrium. This is necessary to understand markets where consumers actually switch because the real costs of switching are shared between firms and those who switch.

800- (toll-free) number portability provides empirical evidence to answer this question. Before portability, a customer had to change numbers to change service providers. This imposed significant switching costs on users, who generally invested heavily to publicize these numbers. In May 1993 a new database made 800-numbers portable. This inter-temporal drop in switching costs and regulations that precluded price discrimination between old and new consumers provide an empirical test of switching costs' effect on price competition.

I use contracts for virtual private network (VPN) services to test how AT&T adjusted its prices for toll-free services in response to portability. Preliminarily (awaiting completion of marginal cost and switching frequency data collection), I find that AT&T reduced margins under VPN contracts that included toll-free usage as the portability date approached, implying that the switching costs due to non-portability made the market less competitive. Contracts which did not include toll-free services and contracts written after portability act as a control group since they were unaffected by portability. These results suggest that, despite toll-free services growing rapidly during this time period, AT&T's incentive to charge a higher price to "locked-in" consumers exceeded its incentive to capture new consumers in the high switching costs era of non-portability.

V. Brian Viard
Graduate School of Business, Stanford University
518 Memorial Way
Stanford, CA 94305-5015
Tel: 650-736-1098
viard_brian@gsb.stanford.edu

This Draft: 10/25/2001

* I would like to thank Dennis W. Carlton, Judith A. Chevalier, Robert Gertner and Fiona Scott Morton for numerous suggestions. I have also benefited greatly from discussions with John Browning, Ann Ducharme, Lars Lefgren, Tomas Serebrisky, Scott Sherburne, Alan D. Viard and Rickard E. Wall. I want to thank George David and Bill Goddard of CCMI, a division of UCG, for their time and generosity in making the tariff data available for my use. Steve Shea of TechCaliber, LLC, Bill Clebsch of Stanford University and Mike Dettorre of Deloitte Consulting contributed enormously to my understanding of the telecommunications industry. I began work on this paper while a PhD student and would like to acknowledge financial support from the University of Chicago Graduate School of Business and the State Farm Companies Foundation. All errors are my own.

Switching costs have important implications for the structure and competitiveness of markets. Even if unable to charge different prices to new and existing consumers due to transactions costs, regulatory constraints or arbitrage possibilities, firms can use aggregate data on purchase history when setting its prices. This can alter the competitiveness of the market and affect the distribution of surplus between consumers and firms. Switching costs are pervasive and can result from a need for compatibility with existing equipment, transaction costs of switching suppliers, costs of learning to use new brands or uncertainty about the quality of untested brands. Klemperer (1995) provides a review of the sources and importance of switching costs. Although economists have developed theoretical switching costs models in which firms charge a single price, these are limited to two-period models or models in which switching costs are assumed to be high enough that no consumers switch in equilibrium. Moreover, few empirical tests of these models exist which is particularly unfortunate given the ambiguous theoretical results. In this paper I address both of these issues.

Previous theoretical work suggests that switching costs have an ambiguous effect on price competition when firms charge the same price to all consumers. This ambiguity is roughly due to two opposing incentives. The firm would like to charge a higher price to previous purchasers who are "locked-in" and a lower price to unattached consumers who offer higher future profitability.[1] Previous work is only suggestive because it is limited to two-period models that suffer from an "end-of-the-world" effect or models that assume switching costs are so high that no consumers switch in equilibrium.

I develop a theoretical model that shows an increase in switching costs may lead to either an increase or decrease in equilibrium prices. Although this model borrows from previous switching costs models, it is the first infinite-horizon model in which consumers actually switch, a necessity for capturing the effects of a change in the level of switching costs. I show that price levels in a market with switching costs relative to a market without such costs depends on the relative number of old and new consumers and the relative importance of "lock-in" versus attracting new consumers as determined by firm and consumer discount factors and the incentive for consumers to switch in the absence of switching costs.

---

[1] It is not necessarily true that the firm would like to charge a lower price to unattached consumers as described below.

It is therefore an empirical question whether these effects are strong enough to reduce price competition. In the second part of this paper I test the effect of switching costs on price competition in the high-growth toll-free telephone service market. Since rapidly growing markets have a greater proportion of new consumers there is a higher probability of switching costs leading to increased price competition. In spite of this rapid growth, I find that switching costs led to lower competition for toll-free services.

800-, or toll-free, service is a telecommunications product in which the receiver, rather than the initiator, pays for the cost of a call.[2] Prior to May 1993, local exchange carriers (LECs), due to a computer database limitation, were unable to route a toll-free call to any inter-exchange carrier (IXC) except that which "owned" the number. Once a customer had contracted with an IXC, they could not change their 800-service to a competing carrier without being assigned a new number. This imposed huge switching costs on firms who used 800-numbers. In May 1993, installation of a new database in all LECs allowed them to route incoming 800- calls to any IXC and allowed consumers of these services to switch IXC providers without changing their phone number. It was necessary for all LECs to implement the database simultaneously since toll-free calls can originate from anywhere in the country. This known, exogenous technological shock lowered switching costs dramatically in May 1993 providing the change in switching costs that I exploit in my empirical tests.[3]

---

[2] The service is often called 800-service because all toll-free numbers originally began with the numbers "800." Toll-free numbers now also begin with "888" and "877."

Due to a regulatory "fairness" doctrine, firms in the 800-services industry had to charge the same price to new and existing consumers. This natural experiment provides an opportunity to test whether changes in switching costs increase or decrease price competition when firms charge the same price to new and old consumers. Consumers enjoyed lower switching costs due to portability. Controlling for other factors, declines in price due to portability would be evidence that switching costs make markets less competitive, while increases in price would be evidence for the opposite.

Using contracts for AT&T virtual private network (VPN) services, I find that portability lowered prices for 800 services implying that higher switching costs under non-portability made the market less competitive. I find that AT&T lowered margins for VPN contracts that contained toll-free services as the portability date approached and that the decrease in margins increased in the intensity of toll-free usage in the contract. I estimate that AT&T lowered its average contribution margin $(p - mc)/p$ on a VPN contract with 50% toll-free usage by 0.282 relative to a contract with no toll-free services for each 100 (expected) days closer portability came. This translates into a decline of about 14% across all contracts in my data set if portability had been implemented before any of the contracts were written.[4] [These results are preliminary awaiting completion of data collection as explained later.]

---

[3] Portability is not completely exogenous if we consider the role of telecommunications firms influencing the Federal Communications Commission (FCC) (the government agency responsible for deciding on portability). If AT&T changed its pricing to influence this decision then there would be a question of causality. AT&T opposed portability so lowering prices with portability, as I find, would not be an obvious method of influence.

[4] As I explain below, the technology for implementing portability was available in 1987 but the court overseeing AT&T's breakup ruled that the technology belonged to AT&T, delaying portability by 6 years.

Overall, my results indicate that AT&T's incentive to charge higher prices to existing consumers subject to the high switching costs of non-portability exceeded its incentive to "lock-in" new users by charging lower prices. Given the rapid growth in 800 services during this time period (AT&T's toll-free minutes were growing over 14% per year), this suggests that switching costs are likely to increase prices in markets with lower growth rates if firms are constrained to charge the same price to new and existing consumers.

Although the primary contribution of this paper is to the switching costs literature, it makes a secondary contribution. A perennial problem in studies of the telecommunications industry has been the difficulty of measuring discounts for services, especially business services. Previous papers have either approximated these discounts or avoided studying business services.[5] I construct a unique data set that fully captures discounts for large users. In the next section I provide background on the toll-free services industry. In Section 2 I review the theoretical and empirical studies of switching costs. Section 3 develops a theoretical model of switching costs. Section 4 describes VPNs and my data and Section 5 the econometric tests I perform. Section 6 discusses the empirical results and I conclude in Section 7.

**1. Toll-Free Services and Portability**

AT&T offered the first interstate toll-free (inbound WATS) service in the United States in 1967. After its divestiture in 1984, other IXCs were legally allowed to provide toll-free service, however the District Court charged with overseeing AT&T's breakup ruled that AT&T retained patent rights over the database technology that allowed LECs to switch toll-free calls to different

---

[5] For example, Knittel (1997) avoids studying business customers: "Residential rates are only used given the higher percentage of businesses that subscribe to discount plans and thus do not pay the retail list rate" (page 529). Even a paper entitled "Competition for 800 Service," by Kaserman and Mayo (1991) contains no actual price data besides a statement that, "For interstate 800 service AT&T has reduced prices by approximately 20% since 1986" (page 405).

IXCs.[6] In 1986, the Federal Communications Commission (FCC) decided, as an interim measure, that toll-free calls would be routed based on the next three digits after 800 (800-NXX-YYYY) referred to as NXX screening. The FCC assigned each IXC one or more NXX prefixes for use in 800-service and the LECs routed all calls beginning with "800-NXX" to the IXC assigned that NXX code. Because of the dependence on NXX, a user who wanted to switch carriers for its toll-free service had to switch numbers.

MCI began offering toll-free service in 1987, followed by Sprint and some smaller IXCs in 1988. Although, users now had a choice of carriers, the NXX screening limitations imposed huge switching costs on toll-free users. Firms usually publish 800-numbers widely, imprinting them on stationary, advertisements and business cards making the cost of changing them significant. A change in numbers also negates any consumer recognition the firm has established and could even harm the firm's reputation if consumers encounter difficulty contacting the firm.[7]

After a lengthy regulatory process, the LECs installed new databases on May 1, 1993, which allowed them to assign and route any 800-call to any IXC. This allowed users to switch providers without changing their phone number. Most popular articles published prior to portability speculated that portability would lower prices for toll-free services.[8] This sentiment has continued in academic articles published since portability. Ward (1993) cites 800-number portability as a reason that long-distance services are more competitive at the writing of his paper than they were in the 1988-1991 period he analyzes. MacAvoy (1995) argues, "That these margins [for inbound WATS service] were increasing rapidly in the latter period when 800 number portability became at least partially available would be counterintuitive in a competitive setting" (page 176). Although academic studies have referred to 800-number portability, none have rigorously analyzed its effect on price competition.

---

[6] The difficulty in switching toll-free calls is that, unlike toll calls, the recipient of the calls pays so that the LEC cannot simply route the call to the initiator's long-distance provider.

[7] For statements in the popular press describing these switching costs see: "Carriers Plot Strategies at Dawn of War Over 800 Users" (*Network World*, November 9, 1992), "Firm Predicts Savings With Tariff 12 Net" (*Network World*, February 12, 1990), "Net Users Remaining Loyal After AT&T's Recent Outage" (*Network World*, January 29, 1990), *Telecommunications Market Sourcebook* (Frost & Sullivan, 1995).

[8] See "Portability Sparks Price Wars" (*Catalog Age*, May 1993), "Airlines + Price Wars = Big 800 Traffic" (*800-900 Review*, Strategic Telemedia, May 1, 1992), "Portability Adds Fuel to 800 Fire" (Karen Burka, *Catalog Age*, October, 1992).

Switching costs can lower prices in a dynamic setting only if the number of new consumers is sufficiently great relative to the number of old consumers. My data indicate that toll-free revenues grew almost nine-fold 1985 to 1999. This measure does not tell us whether new or old consumers generated this growth, but this growth rate is sufficiently high that decreased competitiveness due to switching costs is plausible.

## 2. Switching Costs Literature

Switching costs models that constrain the firm to charge a single price suggest ambiguous results on competition.[9] The most notable two-firm, two-period models, Klemperer (1987a) and (1987b), both reach this conclusion using different assumptions. Klemperer (1987a) assumes differentiated products and motivates switching by assuming a fraction of the consumers experience a change in tastes between the two periods. In this model actual switching occurs. Klemperer (1987b) considers homogeneous products but assumes that consumers differ in the level of switching costs incurred if they switch firms. In this model no consumers switch in equilibrium. In both models switching costs make the second period less competitive than a market without switching costs. In the first period of both models, prices can be either higher or lower than those in a market without switching costs. There are three effects. Firms price lower in the first period because they recognize the value of "locking-in" consumers. Offsetting this are two factors. Consumers anticipate a firm with a lower first period price will charge them a higher price in the second period. This makes consumers' demand less elastic and tends to increase prices. Also, a firm pricing low to build its first period market share invites a more aggressive response from its rival in the second period.

---

[9] There are also switching costs models that consider third-degree price discrimination (see Chen (1997), Nilssen (1992) and Taylor (1999)) and endogenous creation of switching costs (see Caminal and Matutes (1990)). The search costs and network externalities literature are also related.

These two-period models have limited realism. In the first period the firms face demand only from unattached consumers. The second period contains both new and old consumers but an "end-of-the-world" effect distorts the firm's pricing. New consumers in the second period are never valuable as repeat consumers so the firm has no incentive to price lower to capture these consumers. Beggs and Klemperer (1992) model two differentiated-product firms facing new and existing consumers in each period of an infinite-horizon model. Consumers maximize their expected lifetime utility but switching costs are great enough that no one switches in equilibrium. In a symmetric steady-state equilibrium, prices are higher than in a market without switching costs. This result is consistent with that of the second period of the two-period models even though there is no "end-of-the-world" effect. The steady-state assumption, however, is crucial for this unambiguous result. The authors comment that if they add a first period in which neither firm has any old customers, prices in that period are lower than in a market without switching costs. Thus, rapid growth may make it possible for switching costs to lower prices.[10]

Bils (1989) develops a model in which switching costs lead to counter-cyclical markups. He considers a monopolist with an infinite horizon facing overlapping generations of two-period lived consumers. Consumers are uncertain of the product value when young but learn this value perfectly after purchase. This means that old consumers who like the product have less elastic demand for the product than young consumers. This is analytically equivalent to a switching costs model. During a boom the proportion of unattached consumers in the market increases and the firm prices lower to capture new consumers who have more elastic demand. During a downturn, the firm faces proportionately more attached consumers with less elastic demand and prices higher. This result emphasizes the importance of market growth on the relationship between switching costs and price competition.[11]

---

[10] To (1996) extends the Beggs and Klemperer model to focus on switching costs' effect on market shares but maintains the no switching assumption.

In summary, this theoretical work shows that switching costs may either raise or lower prices although the evidence leans toward less competition. Because it is difficult in most contexts to measure switching costs, limited empirical results are available. Borenstein (1991) finds that gasoline stations price discriminated against consumers of leaded gasoline due to the increased switching costs imposed on these consumers as the stations phased it out in favor of unleaded gasoline. Elzinga and Mills (1998), using transaction-level data on wholesale cigarettes, show that customers exhibiting characteristics associated with high switching costs are less likely to switch to a new entrant during a price war. Both of these studies differ from mine in that firms can price discriminate between old and new consumers.

The two papers closest to mine are Knittel (1997) and Sharpe (1997). Knittel finds evidence that rates for long-distance service did not fall after AT&T's divestiture due to search and switching costs, using advertising as a proxy for search costs and the fee charged by local phone companies to change long distance providers as a proxy for switching costs. Sharpe tests the Klemperer (1987a) model result that prices are more competitive the greater consumer turnover in a market. Sharpe finds that the degree of migration into or out of a local market has a positive effect on bank deposit interest rates paid to depositors.

---

[11] Farrell and Shapiro (1988) and Padilla (1995) also consider infinite-horizon switching costs models but they are more difficult to relate to my purposes since they consider an equilibrium in which firms alternate selling to new and old consumers.

## 3. Theoretical Model

The theoretical model I develop in this section serves two purposes. First, it shows that when firms are constrained to charge a single price to all consumers, an increase in switching costs can either increase or lower markups and identifies market conditions under which each of these occurs. This is the first infinite-horizon model to show this. Second, I use the theoretical model as the basis for my econometric model.

My model is the first infinite-horizon, switching costs model in which consumers actually switch. Previous infinite-horizon models assume switching costs are high enough that consumers never switch. In this case, the level of switching costs does not affect prices since all consumers are "locked-in" over the allowable range of switching costs. The authors perform comparative statics by changing the fraction of consumers subject to switching costs. Since portability lowered the *level* of switching costs, it is important for me to consider switching costs over a range that includes the possibility of incomplete "lock-in." I show later in the paper that toll-free customers switched both before and after portability, refuting the possibility of complete "lock-in."

My model extends the two-period model developed by Klemperer (1987a) into an infinite-horizon, overlapping-generations model with two-period lived consumers. I employ a solution technique similar to that in Beggs and Klemperer (1992). I consider two infinitely lived firms whose 800-services are horizontally differentiated. The firms are located at the extremes of a unit Hotelling (1929) line and are symmetric except possibly in their initial market shares. Consumers of 800-services live for two periods and have heterogeneous and uncertain preferences for the two firms' products.[12]

---

[12] Consumers of 800-services are primarily firms but I will to refer to them as consumers to distinguish them from the telecommunications providers (firms).

When young, consumers are uniformly distributed along the line with density one and incur differentiation costs linear in their distance from the firm. For convenience, I normalize the differentiation costs to one.[13] Thus, if a young consumer located at position $x$ on the line purchases from firm A they obtain utility of $r - P_A - x$ where $r$ is the value provided by the product to the consumer located on the firm and $P_A$ is the price charged by firm A. Similarly, if the same consumer purchases from firm B she obtains utility of $r - P_B - (1-x)$ where $P_B$ is the price charged by firm B. The consumer's preferences (or, equivalently, the product features) are uncertain in that, after experiencing a product when young, the utility a consumer obtains from the two products may change. Specifically, a fraction, $\mu$, of consumers are randomly relocated to a new position on the line between the periods in which they are young and old. This reassignment occurs with equal probability for all consumers and is uniform along the line. The remaining fraction, $1-\mu$, experience no change and maintain their original position.

In each time period each firm first sets its price. Consumers then choose their purchases to maximize the net present value of their expected lifetime utility. A young consumer has the option of purchasing from either firm A or firm B and considers the ramifications her decision will have on her options when she is old.[14] An old consumer has the choice of purchasing from the same firm they purchased from when young or switching to the other firm and incurring switching costs of $s$ in addition to the differentiation costs.[15] Between each time period four things happen. First, all old consumers exit the market. Second, a fraction $\rho$ of young consumers, those one period old, also exit the market prematurely. Third, a new generation of young consumers with density one enters the market. Fourth, the uncertainty of preferences for young consumers who remain in the market is resolved.

---

[13] A parameter for differentiation costs only acts as a scale parameter.

[14] I choose $r$ such that the market is covered and not purchasing is sub-optimal.

[15] I again choose $r$ so that the market is covered. Also, I assume consumers incur differentiation costs whenever purchasing, otherwise all old consumers attached to a particular firm would make the same purchase choice.

Each firm is constrained to charge a single price to all consumers in a given period and chooses a sequence of prices to maximize its discounted lifetime profits taking the actions of the other firm as given. The firms' marginal cost is $c$ in each period. I solve for the unique Markov equilibrium in which the firm's customer base is the state variable and the equilibrium price functions are linear. The method of solution is constructive. I first posit the firms' value (profit) and price functions and then solve the consumers' problem to derive the demand function for each firm. Using the demand function I then solve the firms' profit maximization problems by optimizing the Bellman equations. The resulting equations allow me to solve for the unknown constants in the firms' pricing and profit functions.

In solving the model I will focus on firm A since the results for firm B are symmetric. I will let $\sigma_A$ represent the share of old consumers who purchased from firm A last period, $x_A$ the marginal young consumer in the current period, $x_{AB}$ the marginal old consumer in the current period whose realized preferences differ from her ex-ante preferences and bought from A last period and $x_{BA}$ the marginal old consumer in the current period whose realized preferences differ from her ex-ante preferences and bought from B last period.

Suppose that firm A's value and price functions are (where $d, e, k, l, m$ are unknown constants):

(1) $\pi_A(\sigma_A) = k + l\sigma_A + m\sigma_A^2$
(2) $P_A(\sigma_A) = d + e\sigma_A$.

There are five cohorts of demand to consider in each time period: old consumers who purchased from A when young and positions were reassigned with density $(1-\rho)\mu\sigma_A$, old consumers who purchased from B when young and positions were reassigned with density $(1-\rho)\mu\sigma_B = (1-\rho)\mu(1-\sigma_A)$, old consumers whose positions remained the same and purchased from A when young with density $(1-\rho)(1-\mu)\sigma_A$, old consumers whose positions remained the same and purchased from B when young with density $(1-\rho)(1-\mu)\sigma_B = (1-\rho)(1-\mu)(1-\sigma_A)$ and new consumers with density one. I now calculate firm A's demand from each cohort.

The marginal old consumer who purchased from A when young and whose position was reassigned is indifferent between buying from A again and switching to B: $r - P_A - x_{AB} = r - P_B - s - (1 - x_{AB})$ which implies: $x_{AB} = \dfrac{P_B - P_A + 1 + s}{2}$ and demand of $(1-\rho)\mu\sigma_A x_{AB}$. The marginal old consumer who purchased from B when young and position was reassigned is indifferent between switching to A and buying from B again: $r - P_A - s - x_{BA} = r - P_B - (1 - x_{BA})$ which implies: $x_{BA} = \dfrac{P_B - P_A + 1 - s}{2}$ and demand of $(1-\rho)\mu(1-\sigma_A)x_{BA}$. In these two demand equations we see the effect of "lock-in" due to switching costs. Switching costs lower the elasticity of consumers who are part of the firm's customer base and increase the elasticity of those who are not. I will choose parameter values such that all consumers whose preferences remain unchanged purchase from the same firm again (full "lock-in") so that demand is $(1-\rho)(1-\mu)\sigma_A$ from those who purchased from A when young and 0 from those who purchased from B when young.

The marginal new consumer is indifferent between buying from firm A and firm B including the effect it has on their second period utility. In Appendix 1, I show that the position of this marginal consumer is:

(3) $x_A = \frac{1}{2} + b(P_B - P_A)$ where $b = \frac{1}{2(1+\delta_C(1-\rho)(1-\mu+\mu se+(1-\mu)e))}$.

Substituting (2) into (3), I obtain:

(4) $\sigma_A' = \eta - \theta\sigma_A$ where $\sigma_A'$ is next period's market share for firm A, (5) $\eta = \frac{1}{2} + be$ and

(6) $\theta = 2be$.

Using (2), the demand equations derived above and the definition of a value function, I get:

(7) $\pi_A(\sigma_A) = (d + e\sigma_A - c)(\sigma_A' + \mu(1-\rho)(\sigma_A x_{AB} + (1-\sigma_A)x_{BA}) + (1-\mu)(1-\rho)\sigma_A) + \delta_F \pi_A(\sigma_A')$

where $\delta_F$ is the firm discount factor. Note that : (8) $x_{AB} = \frac{1+s+e(1-2\sigma_A)}{2}$ and

(9) $x_{BA} = \frac{1-s+e(1-2\sigma_A)}{2}$.

Firm A chooses its price to maximize its value function taking firm B's choice as given:

(10)

$$\max_{P_A} (P_A - c)[x_A(P_A) + \mu(1-\rho)(\sigma_A x_{AB}(P_A) + (1-\sigma_A)x_{BA}(P_A)) + (1-\mu)(1-\rho)\sigma_A] + \delta_F \pi_A(x_A(P_A))$$

where $x_A$ is as in (3) (before the equilibrium prices are substituted out).

In Appendix 2, I explain how I solve this dynamic programming problem numerically. The problem can only be solved analytically when $\mu = 0$ which is an uninteresting case for my purposes since the switching costs parameter does not influence market prices. I solve for the stable equilibrium ($|\theta| < 1$) that offers the highest profits to the firm.[16] I calculated markups obtained from all combinations of $\delta_C, \delta_F \in \{0.3, 0.5, 0.7\}$, $\rho \in \{0.0, 0.2, ..., 0.8\}$, $\mu \in \{0.1, 0.2, ...0.9\}$ and $s \in \{0.0, 0.1, ..., 1.0\}$ when both firms are in a steady state with equal shares $(\sigma_A = \sigma_B = 0.5)$.[17] Note that for switching costs values that exceed the total differentiation costs of the furthest consumer from the firm (i.e. $s \geq 1$), switching costs no longer affect the markup charged by the firm.[18] This is the case of complete "lock-in" when no consumers find it optimal to switch.

My main result from these simulations is:

> *Result 1: When firms are symmetric and in steady state, an increase in switching costs can either make markets more or less competitive.*

[Insert plot showing the two examples.]. When switching costs increase, several forces are at work on equilibrium prices. It is simplest to consider the effects from the perspective of firm A. First, the higher switching costs allow firm A to charge a higher price to its "locked-in" customer base, both those who preferences remain the same those whose preferences change but not by enough that it is optimal for them to switch to firm B. Second, firm A must offer a lower price to induce consumers whose preferences change to switch from firm B. Third, firm A has an incentive to lower its price to new consumers to build its future customer base. Fourth, because consumers anticipate being "locked-in" once they purchase from a firm they are less tempted by a firm's price cut and their elasticity declines in the level of switching costs. Fifth, higher switching costs increase the importance of inviting a softer response from its rival, providing an increased incentive to price higher. Sixth, those consumers who actually switch bear switching costs. These costs are shared between consumers who switch and the firm based on the relative demand and supply relationship elasticities. Note that in general the switching costs are borne by a fraction of consumers, $(1-\rho)\mu$, ex-post even though ex-ante all consumers face a positive probability of bearing these costs.

---

[16] For some parameter values there are more than one stable equilibrium.

[17] The pricing equation is linear in $c$ so markups are independent of $c$. Output from these calculations is available from the author upon request.

[18] This assumes that $r$ is set so that the consumer at 1 is indifferent between purchasing and not. For higher values of $r$ the firms can sustain higher prices.

Three of these effects act to lower prices while three act to increase them. Which effect is stronger in aggregate depends on the features of the market. My simulations identify the effect of several features on markups.

> *Result 2: When symmetric and in steady-state, firms' markups are decreasing in: a) uncertainty of consumers' preferences and b) firm discount factor and increasing in c) consumer discount factor and d) firm's initial market share. The effect of e) probability that consumers exit the market depends on other parameter values.*

More uncertain preferences leads to a higher probability that the consumer will pay switching costs to obtain their most favored product when old. This decreases the price the consumer is willing to pay when young (part a)). Increasing the firm discount factor (part b)) decreases price because the firm has a greater incentive to build its market share of future "locked-in" consumers. Increasing the consumer discount factor (part c)) has the opposite effect. Consumers are less tempted by a price cut when young, which will "lock" them in when old, if they discount their future utility less (part d)). Part d) follows from the fact that the parameter $e$ in equation 2) is positive in all my simulations (discussed in Appendix 2). A firm with a larger initial market share has more to lose by attracting new consumers through a price cut than a firm with a smaller market share. A higher probability of exit (part a)) can increase or decrease markups depending on the other parameters. A higher probability of exit lowers prices for most parameter because the proportion of "locked-in" consumers is lower. However, for sufficiently high values of discount factors, uncertainty and switching costs, a higher probability of exit can actually increase prices because the firm must bear a sufficient portion of the switching costs that it is better off if more consumers leave the market when old.

The simulations also identify conditions under which switching costs increase competitiveness:

> *Result 3: An increase in switching costs is more likely to result in lower markups when the: a) consumer discount factor is lower, b) firm discount factor is higher, c) product uncertainty is greater, and d) probability of premature exit is greater.*

A decrease in the consumer discount factor widens the regimes in which increased switching costs lowers price. The marginal old consumers earn rents because of the competition between firm A and B. Therefore, although an increase in switching costs lowers consumers' expected second-period utility the firm does not need to adjust the price downward to "make the consumer whole." So the only effect on first-period demand from the increased switching costs is that new consumer demand becomes less elastic (this can be seen in equation 3). A lower consumer discount factor makes the demand elasticity less sensitive to switching costs so an increase in switching costs makes the firm less tempted to cut prices in order to build share for future profits.

An increase in the firm discount factor widens the regimes in which increasing switching costs lowers markups. With a higher discount factor, the firm wants to price lower to enlarge its future "locked-in" customer base. How strong this incentive is depends on the elasticity of young consumer demand. Since this elasticity is declining in switching costs (see equation 3), increased switching costs leads to a greater incentive to lower prices. A higher firm discount factor amplifies this effect so that increased switching costs are more likely to lead to lower markups when the firm discount factor is higher. An increase in product uncertainty increases the range of switching costs values that lower markups because consumers are more likely to face these costs the more uncertain the product features. A higher probability of exit the market also increases this range because a higher proportion of new consumers in the market leads the firm to discount over a wider range of switching costs to "capture" these new consumers.

**4. Virtual Private Network Services Data**

I estimate the effect of portability on margins for AT&T virtual private network (VPN) service. In a VPN an IXC creates a virtual network for medium to large businesses. By specifying ports, corresponding to telephone numbers, within the network and committing to usage volumes, the user receives discounts for calls made to and from these locations, in a manner similar to MCI's *Friends & Family* discount program for residential service in the 1980s. VPNs contain up to five types of voice services, data services and, sometimes, international voice and data services. Three of the voice services are toll services and two are toll-free services. The categories are determined by whether the call utilizes dedicated ("on-net") or switched services ("off-net"). Calls over a dedicated line utilize the LEC's lines but not its switching network, while those over switched lines utilize both. Dedicated service offers lower marginal, but higher fixed, cost than switched service. Firms will choose dedicated service for telephone numbers that initiate or receive high call volumes. Toll calls fall in three categories depending on whether both, one or neither end of the call is "on-net." Toll-free calls fall into two categories depending on whether the call terminates "on-net" or "off-net".[19]

The FCC required IXCs to file tariffs stating rates for all long-distance services including VPNs. The "filed-rate" doctrine of the Communications Act of 1934 (Communications Act) requires all rate-related information to be filed in the tariff upon penalty of $6,000 per offense and $300 per day [47 U.S.C. 203.a].[20] In order to understand how I constructed the data set and why I chose VPN service it is necessary to understand some aspects of the tariff process.

---

[19] Calls to a user's toll-free number originate "off-net" by definition.

The IXCs file two types of tariffs. The first type, baseline tariffs, contains rates available to any user. These tariffs contain volume discounts but do not require the user to pre-commit to a volume level or length of service. The second type, contract-based tariffs, provide discounts off the rates specified in the baseline tariffs for users who commit to certain volume levels, bundles of services, exclusivity arrangements and contract duration. Contract-based tariffs also may contain additional criteria that the carrier must meet in configuring and servicing the more complex networks to which these contracts apply. Baseline tariffs are in effect until the carrier files a subsequent tariff altering the rate, while contract-based tariffs specify a length and are available to any "similarly-situated" customer in the ninety days after its effective date.[21]

Business users could purchase non-VPN toll-free services, but VPN contracts are more convenient for testing the effects of portability for four reasons. First, the largest users of toll-free service, and therefore those most affected by portability, employed VPNs. Second, AT&T began writing contract-based tariffs for VPN services in 1987, well before portability, providing significant data on how AT&T altered its prices in response to portability. For non-VPN services, AT&T did not begin writing contract-based tariffs until early 1992 and changed the baseline tariffs very infrequently. Third, some tariffs included toll-free services ("bundled" contracts) while others did not ("unbundled" contracts). Thus, I can use unbundled contracts, which are unaffected by non-portability, as a benchmark. Fourth, AT&T wrote a significant number of these contracts both before and after portability providing time-series variation.

---

[20] A stronger deterrent for IXCs is their loss of reputation with the FCC.

[21] AT&T offered two types of contract-based tariffs: Tariff 12 options and Contract Tariffs. The FCC required both types to be filed fourteen days before their effective date throughout the time period of my study (except for corrections to a tariff which could be filed three days in advance).

I focus on the interstate market for VPN service because of its relative importance in 800-services.[22] The interstate market is a single national market and includes all calls originating and terminating in different states regardless of whether it is within the same LATA. Under the Communications Act, the FCC regulates the interstate telecommunications market including 800-services. After AT&T's divestiture, the FCC classified it as a dominant carrier and imposed price regulation for some services. The contracts I study were not subject to price regulations but rather were subject to tariff review. The guidelines for this review did not change during the time period of my study. I comment on potential regulatory effects when I discuss my results.

The Communications Act prohibits "unfair" price discrimination, although it allows carriers to charge different prices based on time of day, type of service or other dimensions that the FCC deems reasonable [47 U.S.C. 201.b and 202.a]. This has come to be known as the requirement that IXCs charge the same price to "similarly-situated" customers. Although the definition of "reasonable," and therefore allowable, differences between customers has been defined by debate between the FCC and the carriers and sometimes by courts, the FCC has generally allowed IXCs to tailor prices only by volume purchased, contract length, mix of services and exclusivity clauses. For the class of switching costs models that I wish to test it is only necessary that carriers charged the same price to old and new consumers. I assess the validity of this assumption when I discuss my results.

---

[22] Since AT&T's divestiture in 1984, the telecommunications regulatory structure has defined three types of markets for 800-services: intra-LATA, intrastate (inter-LATA) and interstate (regardless of whether within the same LATA). The United States is divided into 161 LATAs (local access and transportation areas) each with one or more LECs acting as a regulated monopolist for local service. Intra-LATA revenues represented less than five percent of total toll-free revenues in 1995 according to *Telecommunications Market Sourcebook*, Frost & Sullivan, 1995. [Add data for intrastate services.]

My data consists of AT&T prices for VPN service contained in Tariff 12 options filed with the FCC.[23] Each option specifies the charges for a particular combination of usage volume, services mix, contract length, penalty clauses and change fees. Multiple users can, and generally do, sign up for a single option although AT&T does not have to provide the FCC with information about who subscribes to the option. As evidence that multiple users signed up for options, AT&T has filed only 52 new options as of October 2001 since the 151 issued in my data set. I use data from all Tariff 12 options filed by AT&T between February 1990 and November 1993. [I plan to add data from 1994.] This spans most of the time between the FCC's initial decision to implement portability in February 1989 and its implementation. AT&T, along with MCI and Sprint, comprised ninety-one percent of 800-services sold at the time of portability. Unfortunately, MCI did not begin filing contract-based tariffs until 1992 and Sprint until 1995. For my econometric model it is only necessary to assume that the quality of AT&T's VPN service did not change relative to MCI and Sprint's. This is justified based on press reports[24] and by looking at the time trend of baseline rates for VPN services. [Insert time trend of baseline rates.]

Since the FCC does not index tariffs in any meaningful way, I obtain them from CCMI, a division of UCG, which provides pricing information and analysis to help telecommunications users obtain the lowest prices.[25] An observation, $i$, is an original or revised tariff posted by AT&T on effective date $t$. AT&T often revises an existing option rather than issuing a new one.

---

[23] AT&T also sold VPN services through baseline tariffs for customers but I do not include these prices because only smaller users, who are not directly comparable to Tariff 12 users, utilized them. AT&T began selling VPN services through Contract Tariffs toward the end of my sample period (February 1992). [These are directly comparable to Tariff 12 offerings and I plan to add these to my data set.]

[24] From a technical standpoint, the three carriers were equivalent: "AT&T's reliability pitch may be effective, but analysts say its a smoke-screen because MCI and Sprint's 800 service and backup systems are just as reliable." - "AT&T Fights to Protect Its Territory," Gary Strauss, *USA Today*, March 18, 1993.

[25] I am grateful to George David and Bill Goddard for helping me obtain this data.

As with a new option it is available to any "similarly situated customer" for ninety days after issuance. For each tariff option, I recorded the per-minute price for each of the five voice services $(p_{A,i,j}^t \; j=1,2,...5)$ including all discounts, monthly contract dollar value $(r_i)$, contract duration $(h_i)$, fixed monthly fee $(F_i)$ number of data lines of each speed and number of ports of each type (measured, rate option 1 and rate option 2 measured remote ports). For each option I also set *revis*$_i$ to 1 if the contract is revised. AT&T filed 151 active options during the time period of my study. One of these options did not contain any domestic services and five had particularly complicated discounts[26] leaving me with 145 unique options. 94 of these were original filings and 51 were revisions (the original was filed prior to the period of my study). It also filed 42 changes to these options providing 187 observations.

Since the tariffs tailor per-minute voice rates, $p_{A,i,j}^t$, to time of day and distance, I used the rate for the most common call placed, a one-minute daytime call of four hundred miles. Options often contain monthly volume discounts. In applying these discounts I assumed the user consumed 120% of the option's minimum volume commitment. Users avoid falling below the minimum because penalties usually require paying the shortfall and exceeding the minimum by too far since they could have negotiated further volume discounts under a larger contract. Monthly contract dollar value, $r_i$, is the minimum volume commitment specified in the option per month. Contract duration is the minimum time commitment allowed under the option. Since prices were falling during this time period, earlier renewal was favorable but the exit penalties on most options required the user to pay the minimum charge regardless of how many minutes they consumed.[27] For each option, I also recorded the number of ports to calculate the fixed cost of voice lines per month, $c_{v,i}^t$, and data lines of each speed (ranging from 9.6 kbps to 45 mbps) to calculate the fixed cost of data lines per month, $c_{d,i}^t$ including the cost of international data lines. [I am in the process of collecting international voice services data]. These attributes characterize all major attributes of the options. The options also specify compensation for network outages but these vary little across contracts.

---

[26] [I plan to add these four in a future version of the paper.]

[27] On September 30, 1991 the FCC established what it called a "fresh-look" principle that allowed users to terminate Tariff 12 agreements containing 800 services within ninety days of portability without penalty. [Discuss these and adjust.]

Marginal costs for the five voice services in VPN contracts, $c_j^t \ j = 1,2,...5$, include access fees and operational costs. IXCs pay LECs access fees, which are regulated and published in tariffs with the FCC, to complete their calls. Since access fees vary slightly across LECS and the telecommunications operations of VPN users span multiple LECs, I use an average across all LECs published in FCC (1999). Access fees declined during the period of my study as the FCC shifted the cost of the local infrastructure toward monthly fees for residential long-distance. After portability the FCC allowed the LECs to charge a per-query fee to the IXCs for each lookup of an 800-number. Since this fee varied from a low of 0.22 cents to a high of one cent per lookup across LECs and VPN users generally span multiple LECs, I average across the nine major LECs and assume the average length of a toll-free call is 3.6 minutes.[28]

I take estimates of operational costs from court testimony by AT&T in their June 1990 application to provide intrastate toll-free service in California.[29] Operational costs are constant over different output levels until demand exceeds the capacity of the telephone lines. There is significant evidence that the three firms' capacity constraints were not binding during the time period of my study. Huber et. al. (1992, p.321) cites several studies. Another possible capacity constraint is the available supply of toll-free numbers, but the industry did run out of numbers for the 800 prefix until 1996 and in April 1993 still had 60% of the numbers available (FCC, 1999).

---

[28] The nine LECs for which I have data are Ameritech, Bell Atlantic Corp., BellSouth Corp., Nynex, Pacific Telesis Group, Southwestern Bell Corp., US West Inc., GTE Telephone Co. and Southern New England Telephone Co. This data is taken from "Rates May Deter Use of 800 Portability," *Network World*, May 10, 1993, pp. 23, 24 and 34. The estimate of 3.6 minutes average call length is taken from Strategic Telemedia (1996), p. 64.

[29] John Sumpter estimated operational costs for switched toll service to be 1.01 cents, switched toll-free service to be 1.08 cents, dedicated toll service to be 1.30 cents and dedicated toll-free service to be 1.29 cents in testimony on behalf of AT&T to obtain authority to provide intrastate service in California. Application of AT&T Communications of California, Inc. (U 5002 C), June 18, 1990 as reported in MacAvoy (1996).

Each option provides some, but not perfect, information about the proportion, $w_{i,j}$ with $\sum_{j=1}^{5} w_{i,j} = 1$, of the voice services consumed. The details of how I estimate these proportions are in Appendix 3. $\overline{P_{A,i}^t} = \sum_{j=1}^{5} w_j p_{A,i,j}^t$ is the average voice price and $\overline{c^t} = \sum_{j=1}^{5} w_j c_j^t$ is the average voice marginal cost. The total margin on each contract is: $\left(F_i - c_{d,i}^t - c_{v,i}^t + q_i\left(\overline{P_{A,i}^t} - \overline{c^t}\right)\right)$ where $q_i = \dfrac{r_i - F_i}{\overline{P_{A,i}^t}}$ is the average quantity of voice minutes consumed per month. I normalize price of each contract to one so that the margin on each contract is $1 - c^t$ where $c^t = \dfrac{1}{r_i}\left(c_{d,i}^t + c_{v,i}^t + q_i \overline{c^t}\right)$.

Figure 1 provides summary statistics across all the contracts in my data set for the variables used in calculating contract-specific margins. As the figure shows, "off-net" prices and marginal costs are greater than "on-net." Prices for toll-free service are above those for toll service, while marginal costs for toll-free service differ only slightly from those for toll service due to the small database query charges and difference in operating costs. As a result, margins are greater for toll-free than for toll services.

Since a user's decision to contract at time $t$ depends on the cost of switching at the end of the contract, I constructed an expected portability date. Implementation of portability followed a lengthy regulatory process and there was some uncertainty as to the implementation date. Based on accounts in popular magazines and newspapers I constructed an expected portability date that I summarize in Appendix 4. [Comment on fresh look.] I use this to construct a portability dummy variable $dport^t$ set to 1 if portability was implemented at time $t$ and 0 otherwise and an expected time to portability variable: $tport^t = \max\{E[T]-t, 0\}$ where $E[T]$ is the expected portability date. I calculate the fraction of voice usage that is toll-free, $tffrac_i$, and interact it with the time to portability to measure the effect of portability by intensity of toll-free usage.

I collected measures of the market size of toll-free services $(L^t)$ and AT&T's market share $y^t$. Annual toll-free revenue estimates by firm are available from Levinson, et. al. (1990) from 1985 to 1990 and Strategic Telemedia (1997) from 1992 to 1997. To concatenate these two sources, I stacked them in a regression on an AT&T dummy, an MCI dummy, a source (Levinson versus Strategic Telemedia) dummy and an estimate of total long-distance revenues from the FCC (1998).[30] Using the regression results I obtained predicted values of annual toll-free revenues assuming that Strategic Telemedia was the source. To obtain quarterly revenues, I assumed the same seasonality as total long-distance revenues reported in FCC (1998).

I also gathered measures of the fraction of consumers who switched VPN services. I developed two measures: $retain^t$ and $steal^t$. $retain^t$ is the fraction of consumers from period $t-1$ that AT&T retained in period $t$, while $steal^t$ is the fraction of other carriers' consumers that switched to AT&T in the same time period. Using a list of toll-free numbers for the largest consumer-oriented firms contained in *The Smart Consumer's Directory* (pp. 47-82), I determined whether these numbers were provided by AT&T or not in each year based on annual editions of the *AT&T National Toll-Free Directory* (which contained a comprehensive list of AT&T numbers). [This is not yet completed so for now I have approximated these rates based on press reports for pre-portability $retain^t = 0.99$, $steal = 0.02$ and post-portability ($retain = 0.95$, $steal = 0.05$ ].[31] That these numbers are positive rules out the possibility that switching costs other than non-portability were so great that no one switched after portability. If this were the case, portability would have no effect on prices.

---

[30] I also tried including a time trend but found it to be insignificant.

I constructed three other variables thought to affect the cost to IXCs of providing VPN service and the switching costs of VPN users. I measure voice network remoteness, $vremot_i$ as the ratio of ports located remotely (at the customer's premises) to those located at AT&T central offices. This is meant to capture how far-flung the voice network is configured. Data network remoteness, $dremot_i$ is measured similarly as the ratio of access components (connecting customer premises to AT&T central offices) to network components (connecting AT&T central offices). Finally, I include the fraction of costs devoted to international data service, $iremot_i$.

Figure 2 provides summary statistics for the variables in my data set. Contribution margins ranged from 205 to 62% while the fraction of toll-free minutes in contracts ranged from zero to one. AT&T's share of the toll-free market declined from a high of 80% in the first quarter of 1990 to 68% in the last quarter of 1993. The average contract length was 3.6 years and ranged from three to five years, while the average contract size was seven million minutes a year. The average option was written approximately three-quarters of a year before the expected portability date and 75% were written before portability. About 50% of the contracts were revisions to original contracts (the number of unique contracts is higher because my data set begins in February 1990 and many options had already been modified prior to this). The market grew, on average, 81% over the life of a contract (which implies an average annual growth rate of about 18% based on an average contract duration of 3.6 years).

---

[31] Based on "Winds of Change Sweeping Over Cooped-Up 800 World" (*Network World*, May 3, 1993) and "AT&T & MCI Report 'Fresh Look' Results" (*Internet Week*, August 9, 1993).

## 5. Econometric Test

The estimation strategy is to use the Euler equations implied by equation (10) as moment conditions in a GMM estimator (see Hansen and Singleton (1982)). Kim, Kliger and Vale (2001) employ a similar approach to estimate switching costs and probabilities from aggregated data in a panel data set of Norwegian banks. The Euler equations are sufficient conditions for the firm to be at the optimum of a dynamic optimization problem. The intuition of the Euler equations is that, at an optimum, the firm's profits are unchanged by a slight shift in demand from period $t$ to period $t+h$ where $h$ is the number of periods between pricing decisions. I generate the equations by increasing the period $t$ price slightly and decreasing the period $t+h$ price so as to leave all future time periods unaffected. Before generating the Euler equations, I need to modify my theoretical model slightly to apply it to the data. I modify the model to allow for bundling of services within a contract, exogenous market growth, exogenous quantity discounts and multi-period contracts.

The firm now solves the dynamic problem for each contract "type" $i$ every $h$ periods:

$$\max_{P_i^t} \; r_i(P_i^t - c^t)\left[L^t x_{A,i}^t(P_i^t) + \gamma L^{t-1}\left(\sigma_{A,i}^{t-h} x_{AB,i}^t(P_i^t) + (1-\sigma_{A,i}^{t-h})x_{BA,i}^t(P_i^t)\right) + \varpi L^{t-h}\sigma_{A,i}^{t-h}\right] + \delta_F^h \pi_A\left(L^t x_{A,i}^t(P_i^t)\right)$$

where I assume that market size is the same for all contract types and $\gamma = \left(1-(1-\mu)^h\right)(1-\rho)^h$ and $\varpi = (1-\mu)^h(1-\rho)^h$ are the probabilities that an old consumer was relocated or not relocated, respectively, over the contract duration. Now $x_{AB,i}^t = \dfrac{P_{B,i}^t - P_{A,i}^t + s_i^t + v}{2v}$, $x_{BA,i}^t = \dfrac{P_{B,i}^t - P_{A,i}^t - s_i^t + v}{2v}$,

$$x_{A,i}^t = \frac{1}{2} + b_i^{t+h}(P_{B,i}^t - P_{A,i}^t), \quad v = \frac{1-\delta_C \omega}{1-\delta_C(1-\rho)(1-\mu)}, \quad \text{and} \quad b_i^{t+h} = \left[2\left(v(1+\delta_C^h \omega) + e\delta_C^h(\gamma s_i^{t+h}/v + \omega)\right)\right]^{-1}.$$

The pricing equation is now:

(1') $P_{A,i}^t(\sigma_{A,i}^{t-h}, c^t) = d + e\sigma_{A,i}^{t-h} + c^t + \beta_1 h_i + \beta_2 vremot_i + \beta_3 dremot_i + \beta_4 iremot_i +$

$+ \beta_5 tport^t + \beta_6 tport^t * tffrac_i$

This allows price to vary with marginal cost, contract duration and voice, data and international remoteness. It also allows time to portability to affect price over all contracts and differentially for contracts with toll-free service. I expect $e > 0$ but have no prior expectation on the other parameters. The Euler equations implied are:

$$\left(P_{A,i}^t - c^t\right)\left[-L^t b_i^{t+h} - L^{t-h}\gamma/2v\right]dP_{A,i}^t + \left[L^t x_{A,i}^t + \gamma L^{t-h}\left(x_{BA,i}^t + \sigma_{A,i}^{t-h} s_i^t/v\right) + \omega L^{t-h}\sigma_{A,i}^{t-h}\right]dP_{A,i}^t +$$

$$\delta_F^h\left(P_{A,i}^{t+h} - c^{t+h}\right)\left[-L^{t+h}b_i^{t+2h} - L^t \gamma/2v\right]dP_{A,i}^{t+h} + \delta_F^h\left[L^{t+h}x_{A,i}^{t+h} + \gamma L^t\left(x_{BA,i}^{t+h} + \sigma_{A,i}^t s_i^{t+h}/v\right) + \omega L^t \sigma_{A,i}^t\right]dP_{A,i}^{t+h} = 0$$

with $dP_{A,i}^{t+h} = 2eb^{t+h}dP_{A,i}^t$. Using $x_{A,i}^t = \sigma_{A,i}^t$ and $x_{BA,i}^t = \left(e(1-2\sigma_{A,i}^{t-h}) - s_i^t + v\right)/2v$ and defining $g^t = L^t/L^{t-h}$ and $g^{t+h} = L^{t+h}/L^t$ the Euler equation is:

(2') $-\left(g^t b_i^{t+h} + \gamma/2v\right)\left(P_{A,i}^t - c^t\right) + 2eb^{t+h}\delta_F^h g^t\left(g^{t+h}b_i^{t+2h} + \gamma/2v\right)\left(P_{A,i}^{t+h} - c^{t+h}\right) +$

$g^t\left(\sigma_{A,i}^t - 2eb^{t+h}\delta_F^h g^{t+h}\sigma_{A,i}^{t+h}\right) + \gamma/2v\left[(e+v)(1 - 2eb^{t+h}\delta_F^h g^t) - (s_i^t - 2eb^{t+h}\delta_F^h s_i^{t+h} g^t)\right] +$

$\sigma_{A,i}^{t-h}\left(\gamma/v(s_i^t - e) + \omega\right) - 2eb^{t+h}\delta_F^h g^t \sigma_{A,i}^t\left(\gamma/v(s_i^{t+h} - e) + \omega\right) = 0$

The fraction of consumers AT&T "keeps" and "steals" from other IXCs in period $t$ provide additional moment conditions:

(3') $retain^t = \gamma/2v\left(e\left(1-2\sigma_{A,i}^{t-h}\right)+s_i^t+v\right)+\varpi$

(4') $steal^t = \gamma/2v\left(e\left(1-2\sigma_{A,i}^{t-h}\right)-s_i^t+v\right)$

I observe aggregate (new plus old) market share $(y^t)$ common to all types rather than $\sigma_{A,i}^t$ so I solve $g^t y_A^t = g^t \sigma_{A,i}^t + \sigma_{A,i}^{t-h}\left[\gamma/v\left(s_i^t - e\right)+\varpi\right]+\gamma/2v\left(e - s_i^t + v\right)$ for $\sigma_{A,i}^t$. I parameterize switching costs as: $s_i^t = \alpha_1 + \alpha_2 dport_i^t * tffrac_i + \alpha_3 vremot_i + \alpha_4 dremot_i + \alpha_5 iremot_i$ so that switching costs depend on portability (weighted by intensity of toll-free usage), and remoteness of voice, data and international networks. I expect $\alpha_2 > 0$ reflecting higher switching costs under non-portability, $s_i^t > 0\; \forall t,i$. To ensure that $\rho$ and $\mu$ are between zero and one I estimate $m$ in $\mu = \exp(m)/(1+\exp(m))$ and similarly $r$ for $\rho$.

The estimation strategy is to use (1') through (4') as moment conditions in a GMM estimator. The parameters to be estimated are: $\delta = \{\alpha_1 - \alpha_5, \beta_1 - \beta_6, \mu, \rho, d, e\}$ and the data are $X_i^t = \{P_{A,i}^t, c^t, y_A^t, g^t, retain^t, steal^t, vremot_i, dremot_i, iremot_i, tport_i^t, tffrac_i, h_i, dport_i^t\}$. I set $\delta_F$ and $\delta_C$ equal to 0.909. [I plan to estimate these based on equity betas and perform sensitivity analysis.] As recommended by Hansen and Singleton (1982) I use lagged values of the state and exogenous variables as instruments. This is based on fact that the errors in the firm's decisions should be uncorrelated with information available at time $t$.

Portability affects the equilibrium in two ways. First, $\beta_5$ measures the direct effect of portability on prices for contracts that contain no toll-free services. I expect $\beta_5 = 0$. $\beta_6$ measures how portability differentially affects the prices of contracts that contain toll-free services and how much the effect increases in the intensity of toll-free usage. If $\beta_6 > 0$ then portability led to lower margins on VPN contracts and margins were negatively affected as the intensity of toll-free usage was greater. This would be evidence that switching costs made the market less competitive. If $\beta_6 < 0$ then switching costs made the market more competitive. Second, $\alpha_2$ measures the effect of portability on the level of switching costs. These switching costs affect the number of consumers who switch in equilibrium and also the rate of convergence to the steady-state (equal market shares). Therefore, steady-state welfare is affected only by the first effect while both affect welfare inclusive of transitional effects.

The effect of portability on prices and quantities is identified in three ways in my data. First, contracts varied in whether they originated before or after portability (138 of the 187 contracts originated prior to portability). For contracts originated prior to portability, old subscribers to the contract were "locked-in" by non-portability. Second, contracts varied in their intensity of toll-free usage and therefore the degree to which they were affected by portability. Third, the frequency of switching before versus after portability (as measured by *retain* and *steal*) also aids in identification.

## 6. Preliminary Results

The results in this section are preliminary pending completion of international services costs and switching frequencies as described above. Figure 3 provides preliminary results of the GMM estimation. The results indicate that switching costs due to non-portability made the market for toll-free services less competitive. The results show that margins on the average contract without toll-free service increased as the portability date approached, while margins on contracts with toll-free service declined in proportion to the intensity of toll-free usage. On average, AT&T reduced the contribution margin on a contract with 40% toll-free usage (the average in my data set) 0.230 for each 100 (expected) days closer to portability implementation. Figure 4 shows the decline in contribution margin if portability had been implemented at the beginning of my data set under three different scenarios. The first scenario is for the average contract in my data set if the industry had already reached steady-state (equal market shares). In this case, the contribution margin would be about 25% lower. The second scenario averages across all contracts in my data set ignoring transition effects from portability (i.e. not taking into account the effect of portability on AT&T's market shares). In this case, margins would be 14% lower on average. The third averages all contracts including transition effects from portability [needs to be completed]. The technology for implementing portability was available in 1987 when AT&T's rivals first entered but the court charged with the breakup of AT&T ruled that the necessary technology was proprietary to AT&T.

The signs of the coefficients in general are consistent with expectations. The switching cost equation results reflect higher switching costs for contracts with toll-free usage issued under non-portability than for contracts without toll-free usage or those issued after portability. In the pricing equation, AT&T's price is declining in its previous market share consistent with users being subject to switching costs. Users with longer contracts pay lower prices on average, which is reasonable given declining prices during this period. The estimation implies that all users experienced a change in preferences and that 2% of users exited the market annually. The former is clearly too high while the latter is likely too low. These coefficients will be more accurately estimated when I have completed gathering detailed data for the *retain* and *steal* variables.

If AT&T were able to price discriminate between old and new users in their Tariff 12 contracts then my results would be spurious. Although all Tariff 12 options are publicly available and the FCC requires that AT&T makes them available to any "similarly-situated" customer within ninety days of their filing, AT&T could still price discriminate if they tailored the options specifically enough that only a single user qualified. AT&T's ability to do this is limited by the "filed-rate" doctrine. Since all rate-related items must be filed with the FCC, they are used as information in subsequent negotiations. Moreover, tariffs have the weight of law so even if a user signs a private contract with an IXC, a contradicting tariff will take precedence over the private contract in a court dispute.³² Moreover, resellers of 800-services can arbitrage away any price differences across tariffs.

The U.S. Court of Appeals addressed the question of whether AT&T engaged in unlawful price discrimination in its Tariff 12 offerings. In the case, the plaintiff challenged the FCC's finding that AT&T's Tariff 12 filings were non-discriminatory. The Appeals Court agreed with the FCC and concluded that AT&T's Tariff 12 offerings did not violate the Act because they made the rates available to any customer that meets the contract terms.³³

If AT&T was able to tailor Tariff 12 options sufficiently to price discriminate between old and new users then we should see a difference in prices found in new options versus those in revisions to options already filed. New options would be tailored to new users while revisions to options already filed would be targeted at existing users. To test for such price discrimination I estimated my model with each sub-sample. The second and third columns of Figure 3 show the results. I have not yet performed a formal test of structural change but the results are fairly consistent across the two subsets except that switching costs are lower for contracts written before portability relative to those after. This is likely due to the small number of revised options written prior to portability. [I also plan to compare the model estimated with a subset of small contracts and large contracts based on the idea that large users would be easier to target.]

A concern with using portability as a proxy for switching costs is that the FCC subjected AT&T to price regulation on its stand-alone (non-VPN) toll-free services until the portability date. Since AT&T's VPN offerings were subject only to tariff review and this applied both before and after portability, the coincidence of the FCC lifting price regulations and portability raises a concern of confounding effects only if the FCC carried over its treatment of baseline tariff regulation to the Tariff 12 review process and if this regulation was binding in the first place.[34] There is significant evidence that these regulations did not constrain even AT&T's baseline tariff pricing.

---

³² This was reaffirmed in the Supreme Court decision in American Telephone and Telegraph Co. vs. Central Office Telephone, Inc. (108 F.3d 981).

³³ U.S. Court of Appeals Case No. 92-1013 citing from Sea-Land Service Inc. vs. ICC, 738 F. 2$^{nd}$ 1311, 1317, D.C. Cir. 1984.

The design of the regulation gave AT&T more freedom than it appeared. From March 1989 to May 1993 the FCC imposed price-cap regulation on AT&T's toll-free services sold through baseline tariffs. The regulation was applied by baskets and toll-free services were part of Basket 2. AT&T could change its prices within each basket by five percent in either direction of a price cap index set annually by the FCC. The FCC subdivided Basket 2 into four categories. AT&T could change rates for services within some categories by more than five percent as long as the weighted average across all four categories stayed within the allowed range.[35] The FCC initially set the price cap index at AT&T's existing rates and then adjusted them annually for inflation and reduced them by a 2.5 percent "productivity offset" and a 0.5 percent "consumer productivity dividend." AT&T could also submit tariffs that deviated from the price bands subject to FCC scrutiny. Figure 5 shows data assembled by Hall (1993) showing that AT&T's weighted price was well below the price cap index for Basket 2 services during price cap regulation. Lastly, if price regulation had constrained AT&T's pricing we should observe price increases after portability rather than the decreases I find.

---

[34] A consultant I talked to who worked for AT&T as a salesperson of services prior to portability claimed that AT&T was not at all constrained by price caps in filing their tariffs and the FCC rarely challenged tariffs.

[35] Basket 2 included service categories: 1) Readyline 800 (inbound WATS switched), 2) AT&T 800 (classic inbound WATS), 3) Megacom 800 (inbound WATS dedicated) and 4) other 800.

# 7. Conclusion

In this paper I have tested the effect of switching costs in a market in which firms could not price discriminate between new and existing users. I find that the largest firm in the market reduced its margins due to a decline in switching costs implying that switching costs made the market less competitive. Despite rapid growth in the market, the firm's incentive to exploit its existing "locked-in" users was greater than its incentive to "lock-in" new consumers. These results add to a small body of literature providing empirical evidence to a question theoretically unanswerable and important in many different markets.

Taylor, C. R. (1999). Supplier Surfing: Price-Discrimination in Markets with Repeat Purchases. Working paper, May 1999.

To, T. (1996). Multi-Period Competition with Switching Costs: An Overlapping Generations Formulation. *The Journal of Industrial Economics*, 44, 81-87.

Ward, M. (1993). Measurements of Market Power in Long Distance Telecommunications. Bureau of Economics Staff Report, Federal Trade Commission.36

Figure 1    Contract Specifics

| Variable | Mean | Std | Min | Max | n |
|---|---|---|---|---|---|
| On-to-on toll service price ($ per minute) | 0.0621 | 0.00845 | 0.0410 | 0.0992 | 187 |
| Off-to-on toll service price ($ per minute)* | 0.0925 | 0.00916 | 0.0685 | 0.1160 | 187 |
| Off-to-on toll-free service price ($ per minute) | 0.109 | 0.0146 | 0.0786 | 0.1797 | 165 |
| Off-to-off toll service price ($ per minute) | 0.168 | 0.0107 | 0.138 | 0.205 | 186 |
| Off-to-off toll-free service price ($ per minute) | 0.192 | 0.0186 | 0.149 | 0.233 | 154 |
| On-to-on toll marginal cost ($ per minute) | 0.0371 | 0.000391 | 0.0370 | 0.0380 | 187 |
| Off-to-on toll marginal cost ($ per minute)* | 0.0500 | 0.00150 | 0.0478 | 0.0518 | 187 |
| Off-to-on toll-free marginal cost ($ per minute) | 0.0485 | 0.00108 | 0.0470 | 0.0504 | 165 |
| Off-to-off toll marginal cost ($ per minute) | 0.0806 | 0.00311 | 0.0781 | 0.0882 | 186 |
| Off-to-off toll-free marginal cost ($ per minute) | 0.0813 | 0.00288 | 0.0784 | 0.0886 | 154 |
| On-to-on toll service fraction minutes | 0.119 | 0.0295 | 0.0373 | 0.200 | 187 |
| Off-to-on toll service fraction minutes | 0.388 | 0.0959 | 0.121 | 0.650 | 187 |
| Off-to-on toll-free service fraction minutes | 0.315 | 0.123 | 0.000 | 0.394 | 187 |
| Off-to-off toll service fraction minutes | 0.0895 | 0.0221 | 0.0280 | 0.150 | 187 |
| Off-to-off toll-free service fraction minutes | 0.0884 | 0.112 | 0.000 | 0.692 | 187 |
| Average voice price ($ per minute) | 0.109 | 0.0122 | 0.0796 | 0.155 | 187 |
| Average voice marginal cost ($ per minute) | 0.0532 | 0.00406 | 0.0380 | 0.0752 | 187 |
| Average port costs ($1000 per month) | 28.8 | 87.4 | 0.458 | 686 | 187 |
| Average data line costs ($1000 per month) | 265 | 432 | 10.2 | 3493 | 187 |
| Fixed fee ($1000 per month) | 434 | 676 | 17 | 5498 | 187 |
| Contract Size ($million per year) | 14.0 | 16.1 | 13.0 | 111 | 187 |
| Size (million minutes/year) | 7.04 | 8.94 | 0.422 | 72.5 | 187 |

* Applies to off-to-on or on-to-off service

Figure 2      Summary Statistics (n=187)

| Variable | Mean | Std | Min | Max |
|---|---|---|---|---|
| Average contract margin | 0.451 | 0.0675 | 0.201 | 0.621 |
| Market Share | 0.714 | 0.0286 | 0.677 | 0.800 |
| Size (million minutes/year) | 7.04 | 8.94 | 0.422 | 72.5 |
| Duration (years) | 3.63 | 0.861 | 3 | 5 |
| Portability dummy | 0.738 | 0.441 | 0 | 1 |
| Time to Portability (100 days) | 2.72 | 2.12 | 0 | 5.76 |
| Fraction toll-free minutes | 0.404 | 0.148 | 0 | 0.813 |
| Time to Portability * Fraction toll-free minutes | 1.09 | 1.04 | 0 | 3.66 |
| Revision | 0.497 | 0.501 | 0 | 1 |
| Market growth | 1.81 | 0.224 | 1.51 | 2.41 |
| Voice network remoteness | 0.168 | 0.241 | 0 | 1 |
| Data network remoteness | 0.636 | 0.188 | 0 | 1 |
| Fraction international data | 0.00525 | 0.0249 | 0 | 0.210 |
| Keep | 0.980 | 0.0176 | 0.95 | 0.99 |
| Steal | 0.0279 | 0.0132 | 0.020 | 0.050 |

Figure 3      Estimated Parameter Values from GMM Estimation

| Independent Variable | Base Model | New Options | Revised Options |
|---|---|---|---|
| *Switching Costs Equation* | | | |
| Intercept | 0.808 | 1.06 | 2.32 |
|  | (0.0177) | (0.00484) | (0.110) |
| Dport*tffrac | 0.0519 | 0.0627 | -0.419 |
|  | (0.0263) | (0.00966) | (0.046) |
| Voice Network Remoteness | -0.657 | -0.0828 | 0.982 |
|  | (0.0350) | (0.00252) | (0.305) |
| Data Network Remoteness | 0.579 | 0.152 | -1.86 |
|  | (0.0269) | (0.00876) | (0.163) |
| Fraction International Data | -1.22 | 7.25 | -21.7 |
|  | (0.055) | (1.51) | (4.28) |
| *Pricing Equation* | | | |
| Intercept | -0.646 | 0.0603 | -0.410 |
|  | (0.279) | (0.156) | (0.192) |
| Lagged share | 0.347 | 0.373 | 0.372 |
|  | (0.00491) | (0.00183) | (0.00743) |
| Duration | -0.187 | -0.116 | 0.0558 |
|  | (0.0429) | (0.0321) | (0.0103) |
| Tport*tffrac | 0.564 | 0.815 | 0.499 |
|  | (0.0269) | (0.0357) | (0.0201) |
| Tport | -0.187 | -0.208 | -0.192 |
|  | (0.0404) | (0.0182) | (0.00830) |
| Voice Network Remoteness | 0.439 | 0.592 | -0.968 |
|  | (1.12) | (0.275) | (0.174) |
| Data Network Remoteness | 1.63 | 0.638 | 0.802 |
|  | (0.467) | (0.216) | (0.302) |
| Fraction International Data | -4.30 | -18.8 | -1.40 |
|  | (5.47) | (5.19) | (1.36) |
| *Relocation Probability Equation* | | | |
| M | 401 | 639 | 44.3 |
|  | (0.00) | (0.00) | (0.00) |
| *Exit Probability Equation* | | | |
| R | -3.89 | -2.92 | -3.67 |
|  | (0.0825) | (0.0169) | (0.0519) |
| Observations | 187 | 93 | 94 |
| J Statistic (significance level) | 26.7 (0.18) | 17.9 (0.66) | 20.1 (0.51) |

Standard errors are in parentheses. Instruments are: $c_1^t, c_3^t, c_4^t, c_5^t, c_1^{t-h}, c_3^{t-h}, c_4^{t-h}, \sigma_{A,i}^{t-h-1}$

Figure 4     Estimated Effect on Contribution Margins of Implementing Portability in 1989

| Counterfactual | Margin Under Non-Portability | Margin Under Portability | Percentage Change |
|---|---|---|---|
| "Average" Contract in Steady-State | 0.429 | 0.324 | -24.5% |
| All Contracts Ignoring Transition Effects | 0.451 | 0.309 | -14.2% |
| All Contracts Including Transition Effects | | | |

Figure 5     AT&T's Basket 2 Actual Price Index Relative to Price Cap Index

| Year | Price Cap Index | Actual Index |
|---|---|---|
| 1988 | 100.00 | 100.00 |
| 1989 | 96.80 | 97.30 |
| 1990 | 94.00 | 92.80 |
| 1991 | 93.80 | 93.40 |
| 1992 | 94.10 | 92.50 |
| 1993 | 94.10 | 91.40 |

Source: Hall (1993) taken from FCC, "Price Cap Performance Review for AT&T", January 23, 1993.

# Appendix 1 Position of Marginal Young Consumer

The marginal new consumer is indifferent between buying from firm A and firm B including the effect it has on their second period utility:

$$r - P_A - x_A + \delta_C(1-\rho)\left\{\mu\left[\int_0^{x'_{AB}}(r - P'_A - b)db + \int_{x'_{AB}}^{1}(r - P'_B - s - (1-b))db\right] + (1-\mu)[r - P'_A - x_A]\right\} =$$

$$r - P_B - (1 - x_A) + \delta_C(1-\rho)\left\{\mu\left[\int_0^{x'_{BA}}(r - P'_A - s - b)db + \int_{x'_{BA}}^{1}(r - P'_B - (1-b))db\right] + (1-\mu)[r - P'_B - (1 - x_A)]\right\}$$

where $\delta_C$ is the consumers' discount factor and primes indicate next period values. On the left-hand side of the equation is the consumer's discounted expected utility from purchasing from A. The first term is the profit obtained in the current period. The two integrals measure the utility if the consumer's position changes in the next period and are multiplied by $\mu$, the probability that they change. The first integral measures expected utility if the consumer buys from A again while the second integral measures expected utility if they switch to B. The last term on the left-hand side measures the expected utility if the consumers' preferences do not change. The right-hand side is analogous if the consumer purchases from B.

Using $P'_B - P'_A = e(1 - 2x_A)$, $x'_{AB} - x'_{BA} = s$ and $(x'_{AB})^2 - (x'_{BA})^2 = s(1 + P'_B - P'_A) = s(1 + e(1 - 2x_A))$:

$$x_A = \frac{1}{2} + b(P_B - P_A) \text{ where } b = \frac{1}{2(1 + \delta_C(1-\rho)(1 - \mu + \mu se + (1-\mu)e))}.$$

# Appendix 2 Solving the Theoretical Model

Substituting (8) and (9) into (7) and (4) for $\sigma'_A$, I can equate the coefficients in (7) to those in (1) to obtain:

(A1) $k = (d - c)\left(\eta + \mu(1-\rho)\left(\frac{1 - s + e}{2}\right)\right) + \delta_F(k + l\eta + m\eta^2)$

A1

(A2) $l = (d-c)(-\theta + \mu(1-\rho)(s-e) + (1-\mu)(1-\rho)) + e\left(\eta + \mu(1-\rho)\left(\frac{1-s+e}{2}\right)\right) + \delta_F(-\theta l - 2m\eta\theta)$

(A3) $m = e(-\theta + \mu(1-\rho)(s-e) + (1-\mu)(1-\rho)) + \delta_F m\theta^2$

Maximizing (10), the first-order condition is:

(A4) $(P_A - c)\left[\frac{\partial x_A}{\partial P_A} + \mu(1-\rho)\left(\sigma_A \frac{\partial x_{AB}}{\partial P_A} + (1-\sigma_A)\frac{\partial x_{BA}}{\partial P_A}\right)\right] +$

$[x_A + \mu(1-\rho)(\sigma_A x_{AB} + (1-\sigma_A)x_{BA}) + (1-\mu)(1-\rho)\sigma_A] + \delta_F(l + 2mx_A)\frac{\partial x_A}{\partial P_A} = 0$

The second-order condition for the problem is:

(A5) $2\left[\frac{\partial x_A}{\partial P_A} + \mu(1-\rho)\left(\sigma_A \frac{\partial x_{AB}}{\partial P_A} + (1-\sigma_A)\frac{\partial x_{BA}}{\partial P_A}\right)\right] + 2\delta_F m\left(\frac{\partial x_A}{\partial P_A}\right)^2 < 0$

I confirm in my numerical solutions that this holds. Substituting the equilibrium market shares (4) for $x_A$, (8) for $x_{AB}$ and (9) for $x_{BA}$ into the first-order condition yields:

(A6)

$(P_A - c)\left[-b - \frac{1}{2}\mu(1-\rho)\right] + \left[(\eta - \theta\sigma_A) + \mu(1-\rho)\left(\sigma_A s + \frac{1-s+e}{2} - e\sigma_A\right) + (1-\mu)(1-\rho)\sigma_A\right] -$

$\delta_F(l + 2m(\eta - \theta\sigma_A))b = 0$

Solving for $P_A$ and equating the constants to those in (2) yields:

(A7) $d = c + \dfrac{\eta + \mu(1-\rho)\left(\dfrac{1-s+e}{2}\right) - \delta_F b(l + 2m\eta)}{b + \dfrac{1}{2}\mu(1-\rho)}$

A2

$$\text{(A8)} \quad e = \frac{-\theta + \mu(1-\rho)(s-e) + (1-\mu)(1-\rho) + 2\delta_F m\theta b}{b + \frac{1}{2}\mu(1-\rho)}$$

These equations can only be solved analytically when $\mu = 0$ (no uncertainty about preferences) but in this degenerate case the switching costs parameter does not affect equilibrium prices or shares.[1] When $\mu > 0$, I use Mathematica to solve the equations numerically for a given set of parameter values. Regardless of whether the equations are solved analytically or numerically, the procedure is the same. I first solve equations (A8) and (A3) to eliminate $m$ and obtain $e$ as a function of $\theta$. This is a quadratic equation in $e$ which yields up to two roots as a function of $\theta^2$. I then use this result along with (6) to eliminate $e$ and solve for $\theta$ as a function of the other parameters. This yields a cubic equation in $\theta$ which can yield up to three roots for each of the two possible values for $e$. This can produce up to six possible values for $\theta$, however, I choose the stable solution (i.e. $|\theta| \leq 1$)[2] which yields the highest profit for the firms. For the simulations run so far, I have obtained at most two stable solutions and in cases with two solutions the solution with $\theta > 0$ is more profitable for the firms than the solution with $\theta < 0$.

All of the coefficients can then be calculated. Equation (3) yields $b$, (5) yields $\eta$, (A3) yields $b$, (A2) yields $l$, (A7) yields $d$ and (A1) yields $k$. Finally, I check that the necessary constraints on the problem are satisfied (the second order condition is met, individual rationality for each of the marginal consumer types holds and the marginal young consumer prefers to purchase when young rather than waiting to purchase until old).

In all my numerical solutions, $\theta$ has been positive implying that if firms' shares are not equally divided, they converge to one-half in an oscillatory manner. In each period one firm has a dominant "locked-in" share and prices high giving it a smaller share in the next period. This is implied by the fact that $e$ is positive in all my numerical solutions. The smaller share in the next period leads the firm to price lower to build its "locked-in" customer base and so on.

---

[1] My model with $\rho = \mu = 0$ corresponds to To (1996).

[2] Note that if $|\theta| > 1$ the firms' shares diverge.

# Appendix 3 Estimating Distribution of Voice Services by Contract

The usage of the five voice services in each contract is determined by the number of five different types of ports according to the following matrix:

|  |  | Calls To | | |
|---|---|---|---|---|
|  |  | Measured port or rate option 1 measured remote port | Rate option 2 measured remote port | Outside ports |
| Calls From | Measured port or rate option 1 measured remote port | 1 |  | 2 |
|  | Outside ports | 2 |  | 3 |
|  | Port access telephone numbers | 4 | 5 |  |

Let $m$ equal the number of measured ports, $r1$ rate option 1 measured remote ports, $r2$ rate option 2 measured remote ports, $o$ outside ports and $p$ port access telephone numbers. Then based on the above matrix:

$$2(w_{i,1} + w_{i,1}) + w_{i,4} = m + r1$$

$$w_{i,5} = r2$$

$$2(w_{i,2} + w_{i,3}) = o$$

$$w_{i,4} + w_{i,5} = p$$

The tariff provides $m, r1, r2$ but not $o, p$ so we have four equations and seven unknowns. Using the assumptions in MacAvoy (1995, p.107), I assume $w_{i,2} = 3.25 w_{i,1}$ and $w_{i,3} = 0.75 w_{i,1}$. Finally, I impose: $w_{i,4} = w_{i,2}$. I can now solve these seven equations and impose $\sum_{j=1}^{5} w_{i,j} = 1$ to express the weights as fractions.

# Appendix 4 Expected Portability Date Timeline

| Date | Activities Related to Portability | Expected Portability Date |
|------|-----------------------------------|---------------------------|
| < 3/31/89 | Portability not planned. LECs provide NXX-switching for toll-free calls. | N/A |
| 3/31/89 | FCC makes decision to convert database for portability and expects completion in mid-1991.<br><br>"All BOCs are expected to have the new signaling system [SS7 allowing portability] in place sometime in 1991." *Network World*, "The Numbers Game; Advances in Signaling and Switching are Driving the Evolution of 800 Services," June 19, 1989.<br><br>"Full SS7 implementation is not expected until 1991." *Network World*, "Toll-Free Services Market Set for Explosive Growth," July 3, 1989. | 6/30/91 |
| 5/22/90 | FCC announces that 1991 deadline for portability unrealistic, will set new deadline.<br><br>"US Sprint's Canavan said he is hopeful that his company will be able to offer end-to-end services supported by CCS7 [SS7] between some major metropolitan areas next year. 'It will be '92 or '93 before there is significant coverage.'" *Network World*, "Plodding CCS7 Deployment Delays Advanced Services," August 6, 1990. | 15 months in future |
| 8/2/91 | FCC reschedules portability implementation date to March 1, 1993.<br><br>"We require the BOCs and GTE [the LECs] to meet our revised access time standard within eighteen months of the date this order is released." FCC CC Docket No. 86-10, 6 FCC Rcd 5421, "In the Matter of Provision of Access for 800 Service, September 4, 1991. | 3/1/93 |

| Date | Activities Related to Portability | Expected Portability Date |
| --- | --- | --- |
| 11/21/92 | FCC delays implementation date to May 1, 1993.<br><br>"Although we here grant a fifty-seven day extension of the deadline for implementation of data base access, we fully expect LECs and IXCs to continue working diligently . . . the progress that the industry has made thus far in the implementation process must continue in order for our May 1 deadline to be met." FCC CC Docket No. 86-10, 7 FCC Rcd 8616, "In the Matter of Provision of Access for 800 Service," November 20, 1992. | 5/1/93 |
| 5/1/93 | Portability implemented. | 5/1/93 |